\documentclass[12pt]{iopart}

\newcommand {\la} {\langle}\newcommand {\ra} {\rangle}
\newcommand {\beq} {\begin{eqnarray}}
\newcommand {\eeq} {\end{eqnarray}}
\newcommand {\eeqn} [1] {\label{#1} \end{eqnarray}}
\newcommand {\eol} {\nonumber \\}
\newcommand {\ve} [1] {\mbox{\boldmath $#1$}}

\usepackage{graphicx}
\begin{document}

\title{Perey-effect in Continuum-Discretized Coupled-Channel description of $(d,p)$ reactions}
\author{M. G\'omez-Ramos$^1$, N. K. Timofeyuk$^2$ }
\address{$^1$Departamento de FAMN, Universidad de Sevilla, Apartado 1065, 41080 Sevilla, Spain}
\address{$^2$Department of Physics, Faculty of Engineering and Physical Sciences,\\ University of Surrey, Guildford, GU2 7XH, UK}
%\ead{N.Timofeyuk@surrey.ac.uk}
%\date{August 2018}

\begin{abstract}
     The Perey-effect in two-body channels of $(d,p)$ reactions has been known for a long time. It arises when the nonlocal two-body deuteron-target and/or proton-target problem is approximated by a local one,  manifesting itself in a reduction of the scattering channel wave functions in the nuclear interior. However, the $(d,p)$ reaction mechanism requires explicit accounting for three-body dynamics involving the target and the neutron and proton in the deuteron. Treating nonlocality of the nucleon-target interactions within a three-body context requires significant effort and demands going beyond the widely-used adiabatic approximation, which can be done using a continuum-discretized coupled-channel (CDCC) method. However, the inclusion of nonlocal interactions into the  CDCC description of $(d,p)$ reactions has not been developed yet. Here, we point out that, similarly to the two-body nonlocal case, nonlocality in a three-body channel can be accounted for  by introducing the Perey factors. We explain this procedure and present the first CDCC calculations to our knowledge including the Perey-effect.
     
\end{abstract}

\maketitle

\section{Introduction} 

General nuclear many-body theory states that optical potentials should be nonlocal \cite{Feshbach}. Today several groups are investing significant effort in development of these potentials based on modern ab-initio microscopic approaches  \cite{Idi16,Rot17,Idi17,Rot18,Bur19}.
However,   most phenomenological studies employ local optical model to analyse elastic scattering data, providing either individual or global sets of local optical potential parameters. Elastic scattering cross sections depend on the scattering phase shifts only, which means that they are not sensitive to the scattering wave functions in the nuclear interior, so elastic data  cannot distinguish between local and non-local parameterizations. However, when optical potentials are used to calculate non-elastic scattering cross sections  the role of the internal part of the wave function becomes more important and differences arise between the use of local and nonlocal potentials to generate scattering waves in the entrance and/or exit channels.

To account for the effects of nonlocality in non-elastic nuclear reactions, described by models with two-body scattering wave functions,   Perey introduced   a smooth function $f$ of the projectile-target distance $r$ that  multiplies the two-body scattering wave functions  \cite{Perey}. At large $r$, $f(r) \rightarrow 1$ and it gradually decreases to 0.7-0.8 for  small $r$ associated with the nuclear interior. The  introduction of the Perey-factor $f(r)$ is justified by the small range of nonlocality, of the order of 1 fm, which allows for the nonlocal Schr\"odinger equation to be localized using the local energy approximation.  Such an approximation results in a   Schr\"odinger equation  with a local optical potential and a gradient term (or momentum-dependent term). The latter  could easily be removed  using the Perey factorization. Its role in most cases reduces to  decreasing  reaction cross sections that are sensitive to contributions from the nuclear interior.

The Perey factor is included in some reaction codes (DWUCK4 and DWUCK5 \cite{DWUCKs}, TWOFNR \cite{twofnr}) so that many reaction calculations, including $(d,p)$ reactions, reported over the last 50 years, have the Perey effect accounted for. Very recently, the accuracy of the Perey-effect has been investigated for the proton channel of $(d,p)$ and $(p,d)$ reactions \cite{Tit14,Ros15} by comparing it with calculations that used the exact solution for the proton scattering wave obtained from the nonlocal Schr\"odinger equation directly.  
However, the situation with the deuteron channel for this reactions depends on  the reaction model  used. The distorted-wave Born approximation (DWBA) employs deuteron scattering waves calculated from the two-body Schr\"odinger equation and thus the same Perey-factor approximation approach is justified. However, when deuteron breakup is taken into account, the deuteron-channel wave function is obtained from the $A+n+p$ three-body  Schr\"odinger equation, in which case the   nonlocality for optical potentials needs a special treatment. Exact calculations of this three-body equation within the Faddeev method showed a difference between nonlocal calculations and local-equivalent calculations \cite{Del09}. Since the Faddeev approach is not easily adapted for the analysis of $(d,p)$ experiments, an approach to treat the nonlocality in the deuteron channel has been introduced in \cite{Tim13a} within the adiabatic distorted wave approximation (ADWA), which is a popular method for $(d,p)$ data analysis. This method produced a nonlocal deuteron adiabatic model  which can either be solved via localization with the introduction of a local-equivalent potential and the Perey factor \cite{Tim13b,Wal16} or can be solved exactly \cite{Tit16,Bai17}. At low deuteron incident energies the former method can give results accurate to 3$\%$, which is demonstrated in Fig. 2b of Ref. \cite{Bai17} by curves labelled ``Hulten, full" and ``Hulthen, TJ-NLO".

It was recently noticed that ADWA gives rise to dependence of $(d,p)$ cross sections on high-momentum components in the $n$-$p$ motion in a deuteron when nonlocal nucleon optical potentials are used \cite{Bai16}. Exact Faddeev calculations with a wide range of nucleon-nucleon (NN) interactions and the corresponding high $n$-$p$ momentum content revealed a very limited sensitivity to the choice of deuteron model  \cite{Del18}, pointing out that this dependence should  be spurious.  A leading-order local-equivalent continuum-discretized coupled channel (CDCC) method, developed to take nonlocality of optical potentials into account, has also shown reduced sensitivity to high $n$-$p$ momenta \cite{Gom18}. Extension of this local-equivalent CDCC beyond the leading order requires the development of coupled-channel methods with first derivatives both in diagonal and non-diagonal coupling potentials, in which case, the generalization of the Perey factors may not be possible. However, it was shown in \cite{Tim19} that using local-equivalent nucleon potentials with first derivatives in a three-body problem involves a product of Perey factors for neutron and proton. Such a problem was solved for the $^{40}$Ca$(d,p)^{41}$Ca reaction both in an adiabatic approximation and in he Watanabe folding model \cite{Tim19} and   the Perey-effect was shown to affect the cross section by about 9$\%$.

In this paper, we investigate the Perey-effect in CDCC calculations of $(d,p)$ reactions, which has not been explored previously. We expect the Perey-effect in CDCC with local-equivalent potentials to be similar to that in (not-yet-developed) next-to-leading-order  local-equivalent CDCC with fully nonlocal potentials, so the study of the magnitude of this effect with local-equivalent potentials will be a useful antecedent when solving the nonlocal Schr\"odinger equation in a practical way. On the other hand, the  Perey-effect  can modify the information extracted from analysis of experimental $(d,p)$ data within the local CDCC and, therefore, the relevance of this effect must be explored. In section 2 we remind the reader of the formulation of the local-equivalent model for nonlocal potentials, while in section 3 we discuss the CDCC approach to solve three-body $A+n+p$ model  with these potentials. Results of the CDCC calculations for $(d,p)$ reactions on two chosen targets are given in section 4 and conclusions are drawn in section 5.

\section{Nonlocal two-body problem and its localized version}

The nonlocal two-body $A+N$ problem is described by equation,
\beq
(T -E)\Psi(\ve{r}) = -\int d\ve{r}' \, V(\ve{r},\ve{r}') \Psi(\ve{r}').
\eeqn{NLE}
where $\ve{r} = \ve{r}_A - \ve{r}_N$ is the radius-vector 
between $A$ and $N$, $T$ is the corresponding kinetic energy operator and $E$
is the energy of the $N-A$ system in the centre of mass.  An approximate local-equivalent two-body problem is easily constructed if the non-local potential $V_{NA}$  has the   Perey-Buck form \cite{PB},
\beq
V_{NA}(\ve{r},\ve{r}')=H(\ve{r}-\ve{r}')U_{NA}((\ve{r}+\ve{r}')/2), 
\eeqn{PB}
with the  non-locality factor $H$ of range $\beta$
\beq
H(\ve{x}) = \pi^{-3/2} \beta^{-3} e^{-\left(\frac{\ve{x}}{\beta}\right)^2}.
\eeqn{H}
In a leading order the local  potential $U_{loc}^0$  can be obtained as the solution of the transcendental equation \cite{PB}
\beq
U_{loc}^{0}(r)=U_{NA}(r)\exp\left[-\frac{\mu_N\beta^2}{2\hbar^2}(E-U_{loc}^{0}(r))\right],
\eeqn{Uloc0}
where $\mu_N$ is the reduced mass of the $N+A$ system. 
This equation must be corrected for proton scattering,
by reducing  the centre-of-mass energy $E$ in the r.h.s.
of Eq. (\ref{Uloc})  by the local Coulomb interaction $V_{coul}(r)$, which can be represented by a constant, for example, by $\bar{V}_{coul} = -1.08 + 1.35 ((Z-1)/A^{1/3})$ MeV   as given in \cite{Gia76}.

Taking the next step beyond the leading order, the wave function $\Psi$ is found from the following equation derived in \cite{Fie66},
\beq
(T+{\tilde U}_{loc}+ \nabla F\cdot \nabla -E) \Psi =0.
\eeqn{BLO}
%\beq
%(T-E) \Psi = -\left(U_{loc} - \frac{\hbar^2}{2\mu}\frac{\nabla^2f}{f}\right) \Psi + \frac{\hbar^2}{ \mu}\frac{\nabla f}{f} \nabla \Psi.
%\eeqn{BLO}
Here 
\beq
{\tilde U}_{loc} = U_{loc} - \frac{\hbar^2}{2\mu}\frac{\nabla^2f}{f}, 
\eeqn{tUloc}
\beq
U_{loc} = U_{loc}^0 -   \frac{  \beta^2}{16}    (U_{loc}^0)^{\prime \prime} -\frac{\beta^2}{8} \frac{(U_{loc}^0)^{\prime}}{r} - \frac{ \mu\beta^4}{32\hbar^2} \left[\left(  U_{loc}^0\right)^{\prime}\right]^2 \left(1 -  \frac{ \mu\beta^2}{2\hbar^2}   U_{loc}^0\right )^{-1},
\eeqn{Uloc}
\beq
f(r) = \exp \left( \frac{ \mu\beta^2}{4\hbar^2}U_{loc}^0(r) \right),
\eeqn{Perey}
and 
\beq
\nabla F = - \frac{\hbar^2}{ \mu}\frac{\nabla f}{f} .
\eeqn{}
In this case $\Psi(\ve{r}) = f(r) \varphi(\ve{r})$ and $\varphi$ is found from the local Schr\"odinger equation $(T+U_{loc}-E)\varphi=0$. Both $\Psi$ and $\varphi$ are identical in the asymptotic region giving the same scattering cross sections.

The Schr\"odinger equation (\ref{BLO}) does not include spin-orbit interaction. Since our aim is to use ${\tilde U}_{loc}$ in a three-body problem and solve this problem within the CDCC approach and the implementation of the formalism used does not include spin-orbit, we neglect it everywhere below.
%Since at present there is no mechanism to include spin-orbit force within the CDCC we neglect it everywhere below.

\section{Three-body model with first derivatives and CDCC approach}

Let us consider the three-body Schr\"odinger equation for the three-body wave function $\Psi(\ve{R},\ve{r})$  with two-body interactions ${\tilde U}_{nA}$ and  ${\tilde U}_{pA}$ given by Eq.(\ref{tUloc}):
\beq
( T_3 + V_{np}(\ve{r}) &+& {\tilde U}_{nA}(\ve{r}_n)+\nabla_n{F_n(\ve{r}_n)} \cdot \nabla_n  
+V^c_{pA}(r_{p})
\eol 
 &+& {\tilde U}_{pA}(\ve{r}_p)+ \nabla_p{F_p(\ve{r}_p)} \cdot \nabla_p - E )  \Psi(\ve{R},\ve{r})=0,
\eeqn{3bSE1}
where $T_3 $ is the three-body kinetic energy operator, which is a sum of the operators of the $n$-$p$ relative kinetic energy and deuteron kinetic energy in the centre of mass, $\ve{r}_n$ and $\ve{r}_p$ are the coordinate-vectors of the neutron and proton with respect to the target $A$, while $\ve{r} = \ve{r}_n - \ve{r}_p$ and $\ve{R} = (\ve{r}_n+\ve{r}_p)/2$.  The gradients $\nabla_n$ and $\nabla_p$ are  with respect to variables $\ve{r}_n$ and $\ve{r}_p$ respectively. 
Equation (\ref{3bSE1}) could be transformed to a form that does not contain first derivatives of variable $\ve{R}$ \cite{Tim19}, which  is convenient for  expanding the  wave function $\Psi(\ve{R},\ve{r})$ over  the CDCC basis. This is achieved by introducing the representation
\beq
\Psi(\ve{R},\ve{r}) = P_n(\ve{r}_n) P_p(\ve{r}_p) \varphi (\ve{R},\ve{r})
\eeqn{psiPP}
and  requiring that $P_i(r_i) \rightarrow 1$ for $r_i \rightarrow \infty$, which gives for the nucleon Perey factor $P_N$ ($N$ is $n$ for neutrons and $p$ for protons) the first-order differential equation
\beq
 \frac{\nabla_N P_N}{P_N}= \frac{\mu_{dA}}{2\hbar^2} \nabla_N F_N .
\eeqn{peq}
Then $\varphi(\ve{R},\ve{r})$ satisfies the Schr\"odinger equation 
\beq
[T_3 - E + V_{np}(\ve{r}) &+& {\tilde U}^{\rm eff} _{nA}(\ve{r}_n)+{\tilde U}^{\rm eff}_{pA}(\ve{r}_p) 
+V^c_{pA}(r_{p}) \eol
 &+& \Delta U_1 (\ve{r}_n,\ve{r}_p) + \Delta U_2(\ve{r}_n,\ve{r}_p) 
] \varphi(\ve{R},\ve{r})=0
\eeqn{3bSEIa}
with effective $N-A$  potentials,   given by
\beq
{\tilde U}^{\rm eff}_{NA}  = {\tilde U}_{NA}  -\frac{1}{2} \frac{A+1}{A+2} \nabla^2_N F_N+ \left( 1 - \frac{1}{2} \frac{A+1}{A+2} \right) \frac{\mu_{dA}}{2\hbar^2} (\nabla_NF_N)^2, \\
\Delta U_1 (\ve{r}_n,\ve{r}_p) = \frac{2}{A+2}( \nabla_n{F_n}-\nabla_p{F_p} )\cdot \nabla_r, \\
\Delta U_2(\ve{r}_n,\ve{r}_p) =-\left.\frac{\mu^2_{dA}}{4\hbar^2M_A}
 \nabla_nF_n\cdot\nabla_pF_p \right.
\eeqn{}
 and   additional contributions that could be considered as a three-body force since they depend on the positions of both the neutron and the proton at the same time. One of these contributions has $n-p$ velocity-dependence that comes through $\nabla_r$. It is worth noticing that the nature of this three-body contribution is the recoil of target $A$. For infinitely large $A$ it vanishes.
 
 We will use the CDCC expansion \cite{Raw74,Aus87} to solve Eq. (\ref{3bSEIa}): 
 \beq
\varphi(\ve{R},\ve{r}) = \sum_{i=0}^{n_{\max}} \chi_i(\ve{R}) \phi_i   (\ve{r}),
\eeqn{wf1}
where $\phi_0$ is the deuteron bound state wave function $\phi_d$ and $\phi_i$ $ (i\ne0)$ are continuum bins. The channel distorted waves $\chi_i$ are found from the coupled set of differential equations
\beq
 (T_R + U_C(R) &-& E_d) \chi_i(\ve{R}) =
-
 \sum_{i'=0}^{n_{\max}}     U _{ii'}(\ve{R})  \,\chi_{i'}(\ve{R}),
 \eeqn{nleq2}
where the coupling potentials $U _{ii'}(\ve{R})$ are the matrix elements
\beq
U _{ii'}(\ve{R}) = \la \phi_i | {\tilde U}^{\rm eff} _{nA}(\ve{r}_n)+{\tilde U}^{\rm eff}_{pA}(\ve{r}_p) 
+\Delta U_1 (\ve{r}_n,\ve{r}_p) + \Delta U_2(\ve{r}_n,\ve{r}_p)  |\phi_{i'} \ra.
 \eeqn{}
The contribution from the effective $p$-$A$ and $n$-$A$ potentials ${\tilde U}^{\rm eff} _{NA}$ to $U_{ii'}$ is standard and its calculation is built into computer code FRESCO \cite{FRESCO}. The two new terms have been calculated in \cite{Tim19} but only for a different basis expansion, namely over Weinberg states \cite{JT}, in which the first   component only is retained. %Also, the  calculations in \cite{Tim19} accounted for deuteron $s$-wave state  only. 
The CDCC expressions for $U_{ii'}^{(1)}(R)$ and $U_{ii'}^{(2)}(R)$ arising from $\Delta U_1$ and $\Delta U_2$ terms, respectively,  are easily obtained from generalizations of those in \cite{Tim19}:
\beq 
\Delta U_{ii'}^{(1)}(R) &=& 
  \frac{2}{A+2} \int  d\ve{r} \,  \phi_i(r) \left[\frac{r}{2}\left(\frac{F^{\prime}_n(r_n)}{r_n} +\frac{F^{\prime}_p(r_p)}{r_p}\right) \right. 
  \eol
  &+&\,\,\,\,\,\,\,\,\,\, \,\,\,\,\,\,\,\,\,\,\,\,\,\,\,\,\,\,\,\,\,\,\,\,\,\,\,\,\,\,\,\,\,\,\,\,\,\,\,\,R\nu \left. \left(\frac{F^{\prime}_n(r_n)}{r_n} -\frac{F^{\prime}_p(r_p)}{r_p}\right)\right] \phi^{\prime}_{i'}(r), \label{delU1}
\\
\Delta U_{ii'}^{(2)}(R) &=& - \frac{\mu^2_{dA}}{4 \hbar^2 M_A} \int d\ve{r} \,\phi^*_i(\ve{r}) 
\frac{F^{\prime}_n(r_n) F^{\prime}(r_p)}{r_nr_p}  
\left( R^2 - \frac{1}{4}r^2\right) \,\phi_{i'}(\ve{r}) \label{deladpots2}
\eeq
in which $\nu = \cos(\hat{\ve{R},\ve{r}})$, $r_n = \sqrt{R^2+\nu r R + r^2/4}$ and  $r_p = \sqrt{R^2-\nu r R + r^2/4}$. In these equations $\phi_i$ includes $Y_{00}(\ve{r})$.  In all numerical calculations of $U_{ii'}^{(1)}(R)$ and $U_{ii'}^{(2)}(R)$ below we also assumed   $s$-wave functions $\varphi_i$ only. These terms, originating from the target recoil, are already small and employing bins with non-zero angular momentum $l$, which behave as $r^l$ at small $r$, should not change  our conclusions much.

The amplitude of the $A(d,p)B$ reaction is now obtained as
\beq
T_{(d,p)}= \la \psi_B \psi_p \chi(\ve{R}_{p}) | V_{pn} |\psi_A P_n(\ve{r}_n) P_p(\ve{r}_p) \varphi (\ve{R},\ve{r})
\ra, 
\eeqn{Tdp}
where $\ve{R}_p$ is the radius-vector of the proton in the exit channel with respect to the centre-of-mass of the $p+A$ system. Because $V_{np}$
does not involve any internal coordinates of the target $A$, this amplitude contains the overlap function $I_{AB}(\ve{r}_n)= \la \psi_A| \psi_B\ra$. The neutron Perey factor $P_n(\ve{r}_n)$ depends on the same coordinate $\ve{r}_n$. Therefore, to take this Perey factor into account, the overlap function should be   modified. Moreover, due to the short range of the interaction $V_{np}$ only those values of $\ve{r}_p$ which are very close to $\ve{r}_n$ will contribute to the amplitude $(\ref{Tdp})$. To check this assumption we have performed a series of simple finite- and zero-range ADWA calculations for all those cases  considered below and found that finite-range effects in the first maximum do not exceed 2.5$\%$. Thus, to take the Perey-effect into account it is sufficient  to use a modified overlap function $I_{AB}^{\rm mod} (\ve{r}_n)=P_n(r_n)P_p(r_n) I_{AB}(\ve{r}_n)$ in a standard CDCC calculation  (either zero-range or finite-range) with effective nucleon optical potentials ${\tilde U}^{\rm eff}_{NA}$ supplemented by  three-body  recoil terms $\Delta U_1$ and $\Delta U_2$.  Below, all the CDCC calculations treated finite-range exactly.

\section{Perey-effect study for nonlocal optical potentials}

We have performed numerical calculations for the $^{12}$C($d,p)^{13}$C reaction populating the lowest 1/2$^-$, $1/2^+$ and $5/2^+$ states at $E_d = 30$ MeV and for $^{40}$Ca($d,p)^{41}$Ca(g.s.) reaction at three incident deuteron energies $E_d= 10$, 20 and 56 MeV. The non-local energy-independent optical potential $U_{NA}$ from \cite{Gia76} was used to construct $U^{0}_{loc}$, ${\tilde U}^{\rm eff}_{NA}$, $\nabla F_{N}$ and $P_N$  in the entrance deuteron channel and the same parameterization was used to construct  optical potentials (\ref{tUloc}) and (\ref{Uloc}) in the exit proton channel.  In \cite{Gia76} the nonlocality range is   $\beta = 0.9552 \left(\frac{A+1}{A}\right)^{1/2}$ fm and  the $U_{NA}$ is given by a real part Woods-Saxon form with the depth of  88.6 MeV, radius of $1.25A^{1/3} - 0.282$ fm and diffuseness $a$ of 0.57 fm. Imaginary potential is of the surface type with the depth of 23.7 MeV and with the same radius and diffuseness as for the real part. The spin-orbit potential in neglected  in the present work.
The standard Perey factors (\ref{Perey}) were applied in the outgoing proton channels as well.
In the case of $^{12}$C target the imaginary part of the nonlocal potential from \cite{Gia76} was reduced to 13 MeV according to \cite{Del09} to prevent  unphysically large absorption. We have also used the overlap between $^{12}$C and $^{13}$C represented by a single-particle wave function obtained in the Wood-Saxon potential well of radius $1.1359A^{1/3}$ fm and diffuseness $0.57$ fm as suggested in \cite{Del09}. The spin-orbit potential's geometry was the same  and its depth was 5.5 MeV. For $^{40}$Ca we adopted the overlap function from \cite{Wal16} where the radius $ 1.252 A^{1/3}$ rm and diffuseness 0.718 fm were used both for the central and spin-orbit Woods-Saxon potential and the depth of the spin-orbit potential was 6.25 MeV.  In both cases the depth of the central potential was adjusted to reproduce the bound neutron separation energy. The spectroscopic factors were kept equal to one everywhere.

The finite-range CDCC calculations were done using the code FRESCO \cite{FRESCO} considering only the $s$-wave continuum which was discretized in eight bins equispaced in momentum with an energy from 0 to 24 MeV, for the $^{12}$C($d,p)^{13}$C reaction, 7 bins from 0 to 7 MeV for the $^{40}$Ca($d,p)^{41}$Ca reaction at 10 MeV, 8 bins from 0 to 16 MeV for the reaction at 20 MeV and 10 bins between 0 and 50 MeV for the reaction at 56 MeV. In these calculations we used the Hulth\'en model of the $n$-$p$ interaction \cite{Hulthen}. This potential gives the same low-momentum $n$-$p$ behaviour as all the modern NN models do \cite{Bai16} and the dominance of the low $n$-$p$ momentum contribution to the $(d,p)$ reactions has been confirmed by rigorous nonlocal Faddeev calculations in \cite{Del18}. We should note that derivatives in Eqs. (\ref{delU1})-(\ref{deladpots2}) may suggest an enhanced sensitivity to the $n$-$p$ interaction model, however, the contributions $\Delta U^{(1)}_{ii'}$ and $\Delta U^{(2)}_{ii'}$ are recoil-induced and enter the CDCC equations with a scaling factor of $1/A$ thus weakening a possible $n$-$p$ model dependence. In these calculations we neglected contribution from the $d$-wave continuum. Test calculations for $^{40}$Ca at $E_d = 20$ MeV showed that this contribution does not exceed 3\% at the maximum of the angular distributions.

%\begin{figure}[t]
%centering
%\includegraphics[scale=0.32]{temp-0eff.eps}
%\includegraphics[scale=0.32]{temp-eff12.eps}
%\includegraphics[scale=0.32]{temp-all2P.eps}
%\includegraphics[scale=0.32]{temp-eff2nlcdcc.eps}
%\includegraphics[scale=0.32]{JPhysG-ca40-0eff.eps}
%\includegraphics[scale=0.32]{JPhysG-ca40-effall.eps}
%\caption{Cross sections ratio $\sigma_0/\sigma_{\rm eff}$ ($a$) and  $\sigma_{\rm eff}/\sigma_{\rm all}$ ($b$) for $^{12}$C($d,p)^{13}$C reaction populating the first lowest $1/2^-$, $1/2^+$ and $5/2^+$ final states. See text for further explanations.
%}
%\label{fig:ratio:c12}
%\end{figure}

\begin{figure}[t]
\centering
\includegraphics[width=0.9\textwidth]{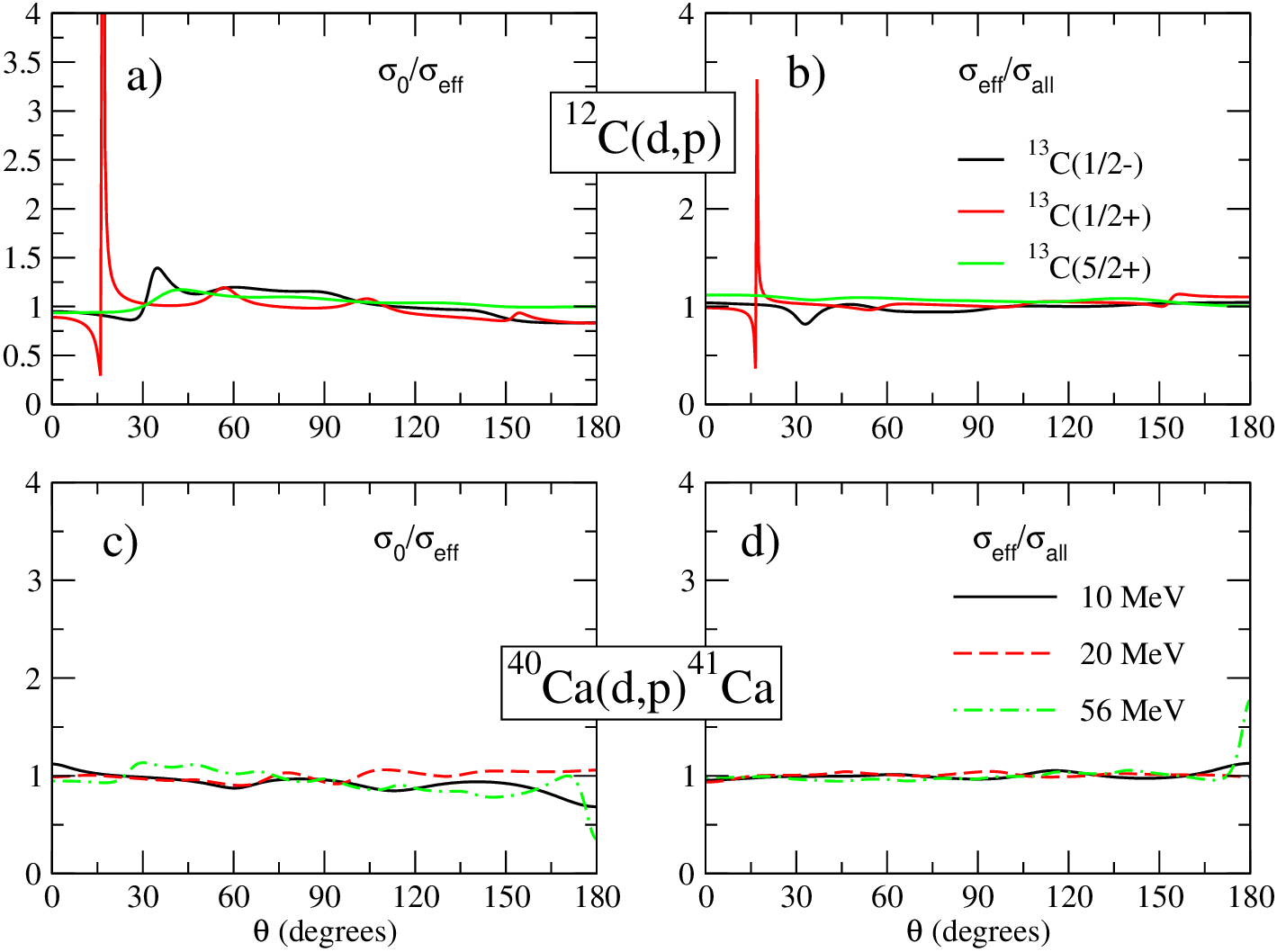}
\caption{Cross sections ratio $\sigma_0/\sigma_{\rm eff}$ ($a$) and  $\sigma_{\rm eff}/\sigma_{\rm all}$ ($b$) for $^{12}$C($d,p)^{13}$C reaction, and for the $^{40}$Ca($d,p)^{41}$Ca reaction, c) and d) respectively. For the $^{12}$C($d,p)^{13}$C reaction, results are shown for the population of the  $1/2^-$, $1/2^+$ and $5/2^+$ final states, while for the $^{40}$Ca($d,p)^{41}$Ca, results are shown for three different reaction energies: 10, 20 and 56 MeV.  See text for further explanations.
}
\label{fig:c12ca40}
\end{figure}

We have compared several CDCC calculations for $(d,p)$ cross sections without Perey-factors. First of all, differential cross sections $\sigma_0$ were calculated using the leading-order local-equivalent neutron and proton potentials $U_{loc}^0$ only. Then we calculated differential cross sections $\sigma_{\rm eff}$ with the modified next-to-leading order two-body nucleon potentials ${\tilde U}^{\rm eff}_{NA}$. Finally, we added recoil-induced three-body terms $\Delta U_1$ and $\Delta U_2$ and obtained the differential cross sections $\sigma_{\rm all}$.  We found that it is more informative to plot ratios between  $\sigma_0$, $\sigma_{\rm eff}$ and $ \sigma_{\rm all}$ as all of them look rather similar. %Fig. \ref{fig:ratio:c12} shows $\sigma_0/\sigma_{\rm eff}$ and $\sigma_{\rm eff}/\sigma_{\rm all}$ for $^{12}$C$(d,p)^{13}$C reaction populating the first lowest $1/2^-$, $1/2^+$ and $5/2^+$ final states while Fig. \ref{fig:ratio:ca40} shows the same quantities for $^{40}$Ca$(d,p)^{41}$Ca for three deuteron incident energies: 10, 20 and 56 MeV.   
Results are shown in Fig.~\ref{fig:c12ca40}. The top panels correspond to the ratios $\sigma_0/\sigma_{\rm eff}$ a) and $\sigma_{\rm eff}/\sigma_{\rm all}$ b) for $^{12}$C$(d,p)^{13}$C reaction populating the first lowest $1/2^-$, $1/2^+$ and $5/2^+$ final states while the bottom panels c), d) show the same quantities for $^{40}$Ca$(d,p)^{41}$Ca for three deuteron incident energies: 10, 20 and 56 MeV. 
The values of these ratios in the first peak are also given in Table \ref{tab:ratios}. It can be seen that the modification of $U^0_{loc}$ can affect $\sigma_0$ by up to 7$\%$ if the final neutron bound state does not have a node. For the population of the final $1s_{1/2}$ state in $^{13}$C this effect may be stronger, up to 11$\%$. Adding recoil-induced three-body force in most cases has less effect and it is much smaller for the heavier target $^{40}$Ca, as expected. We would also like to note that in a particular case of populating the weekly-bound $s$-wave  $^{13}$C(1/2$^+$) state the differential cross sections decrease fast with the scattering angle. Adding small corrections to $U_{loc}^0$ changes the radius of the local optical potentials in the CDCC equations and, therefore,  slightly shifts the position of the  first minimum of the angular distribution thus creating a sharp jump of the ratio $\sigma_o/\sigma_{\rm eff}$ in the vicinity of this minimum. This jump therefore does not carry any important information.

%\begin{figure}[b]
%\centering
%\includegraphics[scale=0.32]{temp-0eff.eps}
%\includegraphics[scale=0.32]{temp-eff12.eps}
%\includegraphics[scale=0.32]{temp-all2P.eps}
%\includegraphics[scale=0.32]{temp-eff2nlcdcc.eps}
%\includegraphics[scale=0.32]{JPhysG-ca40-0eff.eps}
%\includegraphics[scale=0.32]{JPhysG-ca40-effall.eps}
%\caption{The same as Fig. \ref{fig:ratio:c12} but for $^{40}$Ca$(d,p)^{41}$Ca reaction for deuteron incident energies of 10, 20 and 56 MeV.
%}
%\label{fig:ratio:ca40}
%\end{figure}

The role of the Perey factor is demonstrated in Fig.  \ref{fig:xsecs} for the same reactions. Only for one case, $^{40}$Ca$(d,p)^{41}$Ca at $E_d = 10$ MeV, does the introduction of the Perey-factor  not affect the cross sections in the main maximum, which is most likely caused by the more peripheral nature of this reaction due to its low incident energy. Indeed, as discussed in section 3, including the Perey factor can be applied to  the overlap function, in other words, the Perey-effect reduces this function in the nuclear interior by  70-80$\%$. Insensitivity of the peak cross section  to this reduction is a sign of the peripherality of this nuclear reaction.    In all other cases, the cross section at the first peak is affected to a significant factor, displayed in Table \ref{tab:ratios}, showing that the internal contributions become more important. For those cases where the overlap function of the transferred nucleon does not have any nodes the peak cross section decreases by 7-19$\%$. For the transfer of a neutron to the $1s_{1/2}$ state with one node the Perey factor redistributes the cross section between the first and the second maxima, leading to differences of 23$\%$ and 47$\%$, respectively. In general, the cross sections outside the fist maxima seem to be affected  by the Perey effect more strongly.

\begin{table}[t]
\caption{Ratios of various first-peak differential cross sections for two reactions,$^{12}$C$(d,p)^{13}$C and $^{40}$Ca$(d,p)^{41}$Ca. For $^{12}$C$(d,p)^{13}$C($1/2^+$), these ratios are  shown at the second peak as well. The angle at which the ratios have been calculated are shown in the third column. See text for detailed notation.
}
\centering
\begin{tabular} {p {4 cm} p{ 1.  cm}  p{ 1.  cm} 
%p{ 1.7 cm} p{ 1.7 cm} p{ 1.7 cm} 
ccc
c  }  
\hline
reaction & $E_d$ & $\theta_{\rm peak}$ & $\sigma_0/\sigma_{\rm eff}$ & $\sigma_{\rm eff}/\sigma_{\rm all}$ & $\sigma_{\rm all}/\sigma_{\rm all}^{\rm P}$ & $\sigma_{\rm all}/\sigma_{\rm LECDCC}$\\
\hline
\hline
$^{12}$C$(d,p)^{13}$C($1/2^-$) & 30 & 6 & 0.942 & 1.03 & 1.15 & 0.978 \\
$^{12}$C$(d,p)^{13}$C($1/2^+$) & 30 & 0 & 0.89 & 0.984 & 0.77 & 1.046\\
  &   & 32 & 1.02 & 1.03 & 1.47 & 0.782\\
  $^{12}$C$(d,p)^{13}$C($5/2^+$) & 30 & 5 & 0.936 & 1.12 & 1.19 & 0.986\\
$^{40}$Ca$(d,p)^{41}$Ca($7/2^-$) & 10 & 37 & 0.973 & 0.995 & 1.005 & 0.93 \\
$^{40}$Ca$(d,p)^{41}$Ca($7/2^-$) & 20 & 25.5 & 0.979 & 1.006 & 1.07 & 0.93 \\
$^{40}$Ca$(d,p)^{41}$Ca($7/2^-$) & 56 & 0 & 0.946 & 0.987 & 1.34 & 1.21 \\
\hline\hline
\end{tabular}
\label{tab:ratios}
\end{table}

\section{Comparison to Local-Equivalent CDCC}

So far we considered the CDCC method based on nucleon optical potentials in which non-locality is contained in the first-order derivatives only. Such potentials arise in the process of localization of the nonlocal two-body problem. No methods have yet been developed  to solve CDCC equations for nonlocal optical potentials. However, a leading-order local-equivalent CDCC - LECDCC - has been derived in \cite{Gom18} for nonlocal potentials of the Perey-Buck type (\ref{PB}). We can now compare the CDCC with local potentials $U^0_{loc}$ with the LECDCC results for the same underlying nonlocal potential.

In the LECDCC,  with the assumption of just the $s$-wave continuum of the deuteron, the channel functions $\chi_i(R)$ are computed from a coupled set of differential equations (\ref{nleq2})
%\beq
%(T_R + U_C(R) &-& E_d) \chi_i(\ve{R}) = -\sum_{i'} U_{ii'}^{\rm loc} (\ve{R}) \chi_{i'}(\ve{R}), \,\,\,\,\,\,\,\,\,
%\eeqn{LECDCC}
with local-equivalent coupling potentials $U_{ii'}^{\rm loc}$ that satisfy
a system of the transcendental matrix equations written for
%\beq
$X_{ii'} \equiv (E_{i'}-U_C) \delta_{ii'} - U^{\rm loc}_{ii'}$:
%\eeqn{nlineq1}
\beq
f_{ii'}^{(0)} - (E_j-U_C) \delta_{ij} &+& \sum_k (f_{ii'}^{(1)} +\delta_{ik})X_{ki'}
+ \sum_{kl}f_{ii'}^{(2)}X_{kl}X_{li'} +...=0, \,\,\,\,\,\,\,\,\,\,\,\,\,\,\,
\eeqn{nl1}
where %$f_{ij}^{(n)}= \gamma_n  U^{(n)}_{ii'}$,
 \beq
%  \gamma_n 
  f_{ij}^{(n)}= \frac{(-)^n}{n!(2n+1)!!}
 \left(\frac{\mu_d \alpha_2^2 \beta^2}{4\hbar^2} \right)^n U^{(n)}_{ii'}
 \eeqn{}
and %the coupling potential
\beq
 U^{(n)}_{ii'}(\ve{R}) =  \int d\ve{x}\left[{\bar \phi}_i^{(n)}(\ve{x}) \right]^{*} \left[ \sum_{N} U_{NA}\left( \frac{\ve{x}}{2}-\ve{R}\right)\right] \phi_{i'}(\ve{x}). 
 \eeqn{}
 The coupling potentials $U^{(n)}_{ii'}$ are constructed using the usual bin functions $\phi_i$ and the continuum bin functions modified by nonlocality
 \beq
 {\bar \phi}_i^{(n)}(\ve{x}) = \int d\ve{s}\, H(s) \left(\frac{s}{\beta}\right)^{2n} %\frac{s^{2n}}{\beta^{2n}} \,
 \phi_i(\ve{x}+\alpha_1\ve{s}).
 \eeqn{modbins}
Transcendental equations (\ref{nl1}) have been solved in \cite{Gom18} using the Newton method, with the obtained $U_{ii'}^{\rm loc}$ being read into the CDCC reaction code FRESCO.
Although the left-hand side of (\ref{nl1}) contains an infinite sum of terms it was found that truncating it up to cubic terms on $X_{il}$  was sufficient to get  converged solutions $U_{ii'}^{\rm loc}$  with a good accuracy. The difference between the cubic and quadratic approximations
was less than 0.5$\%$ for the values of $R$ most important for reaction calculations.
%was about 4.5$\%$ in the nuclear interior decreasing to 0.1$\%$  at 4.5 fm. 
%(less than 5\% change in the potentials).

\begin{figure}[t]
\centering
\includegraphics[scale=0.30]{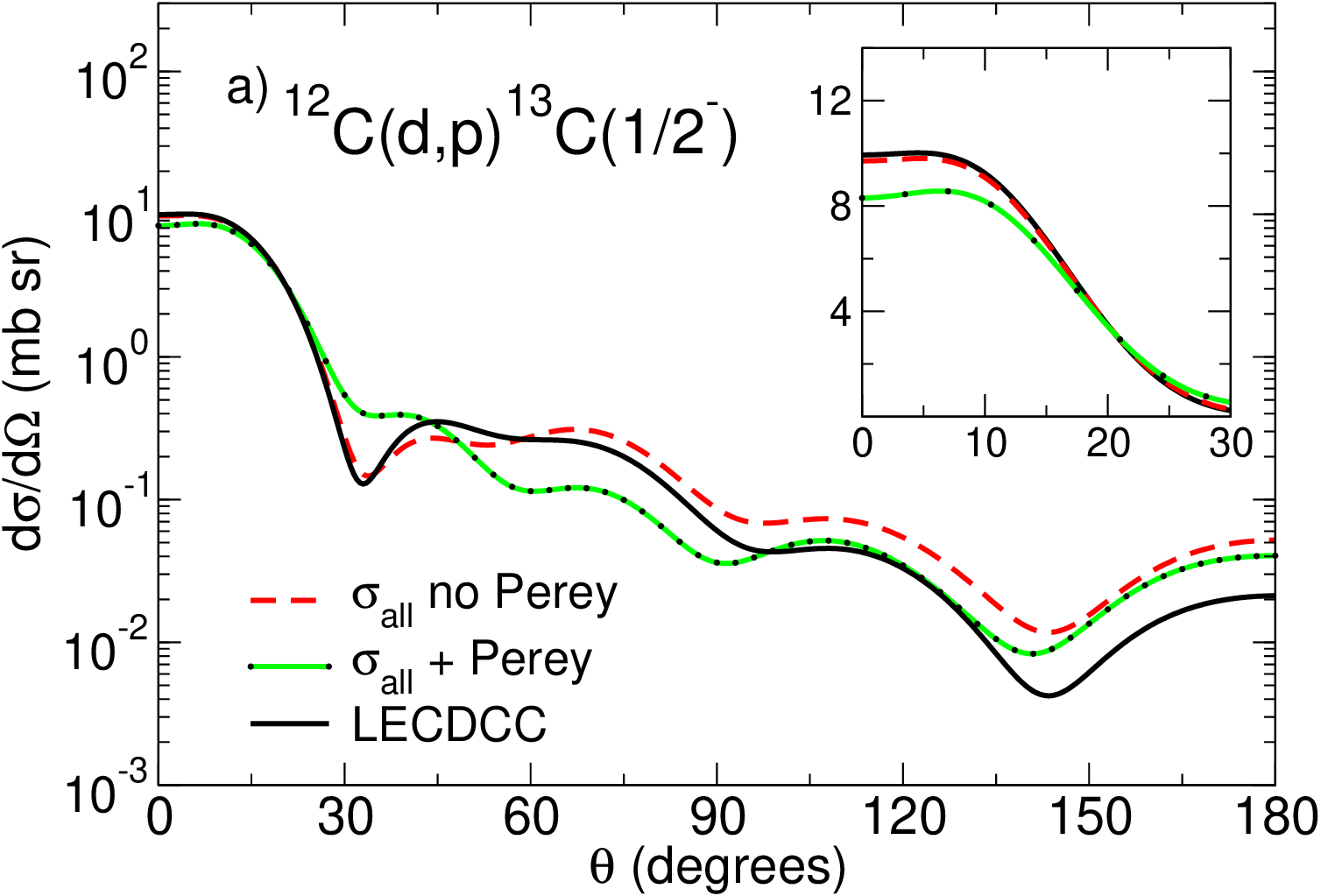}
\includegraphics[scale=0.30]{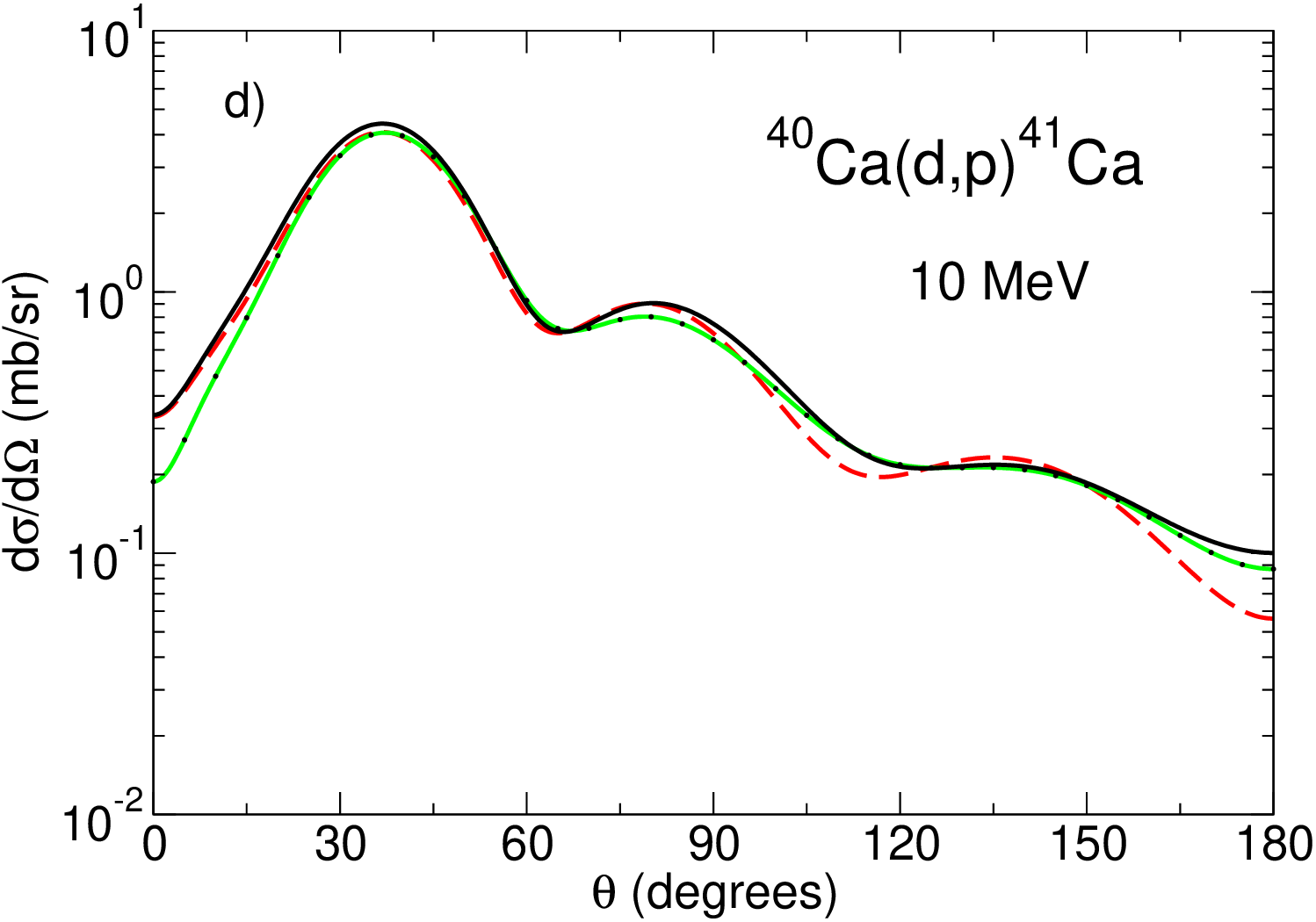}
\includegraphics[scale=0.30]{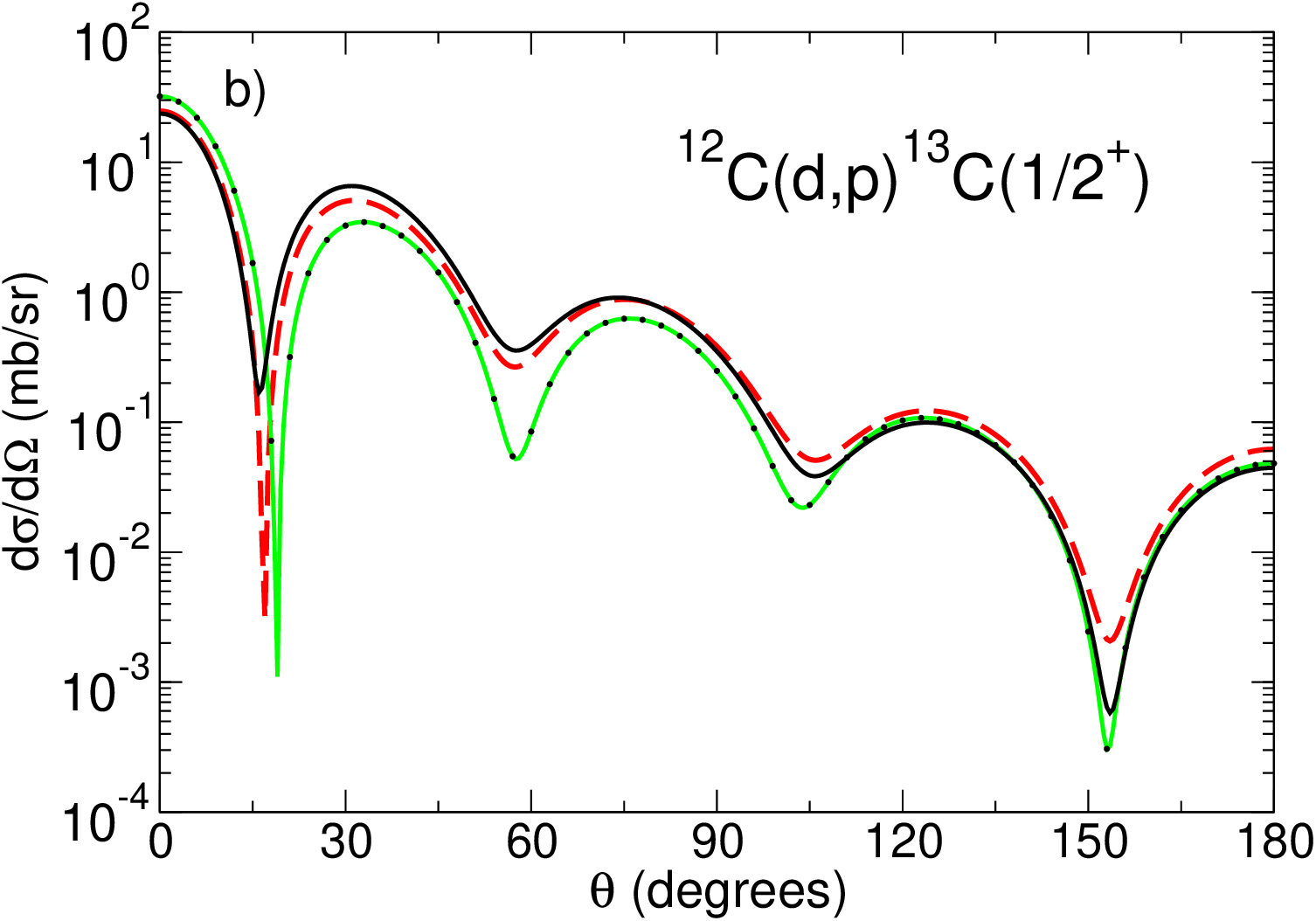}
\includegraphics[scale=0.30]{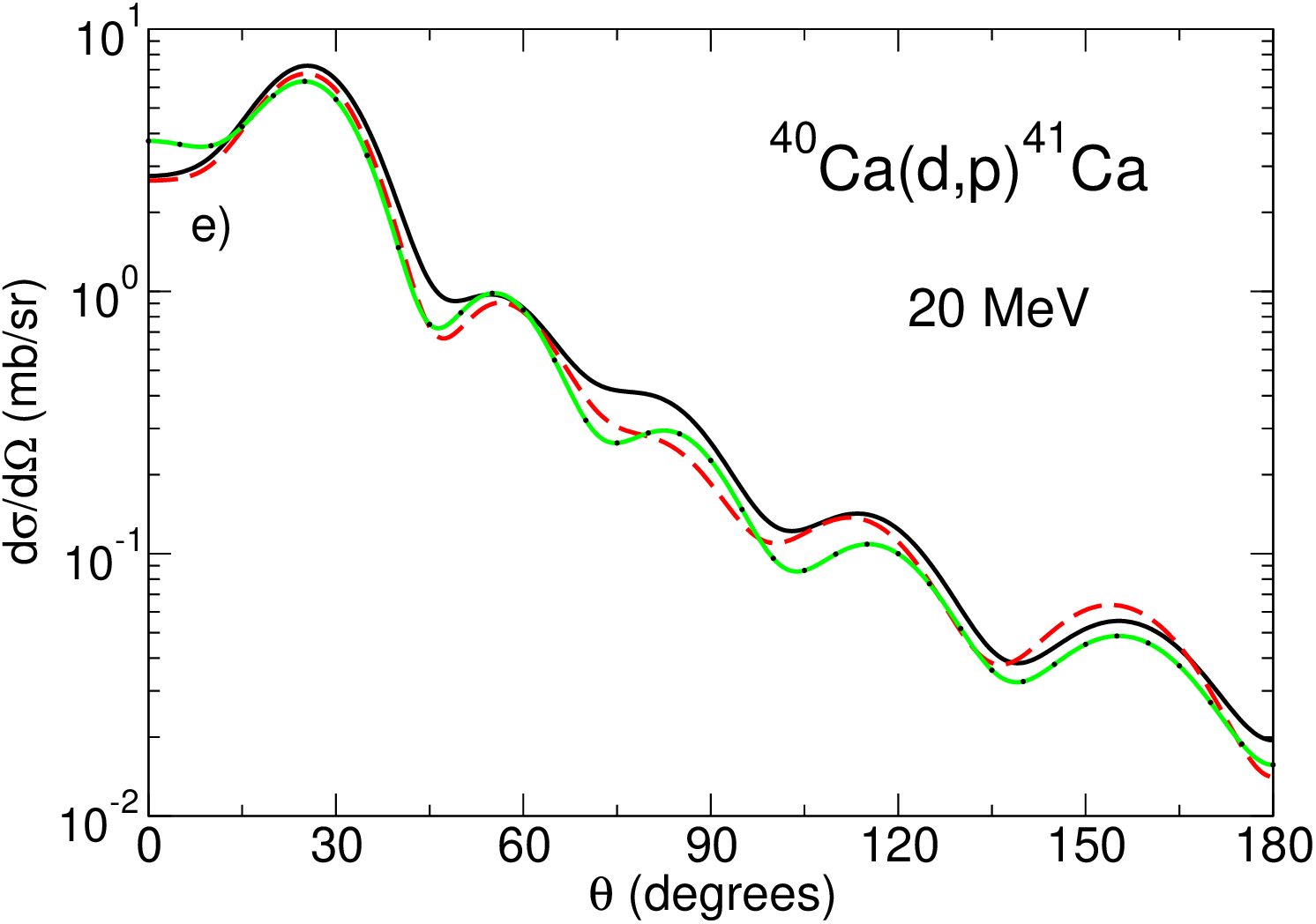}
\includegraphics[scale=0.30]{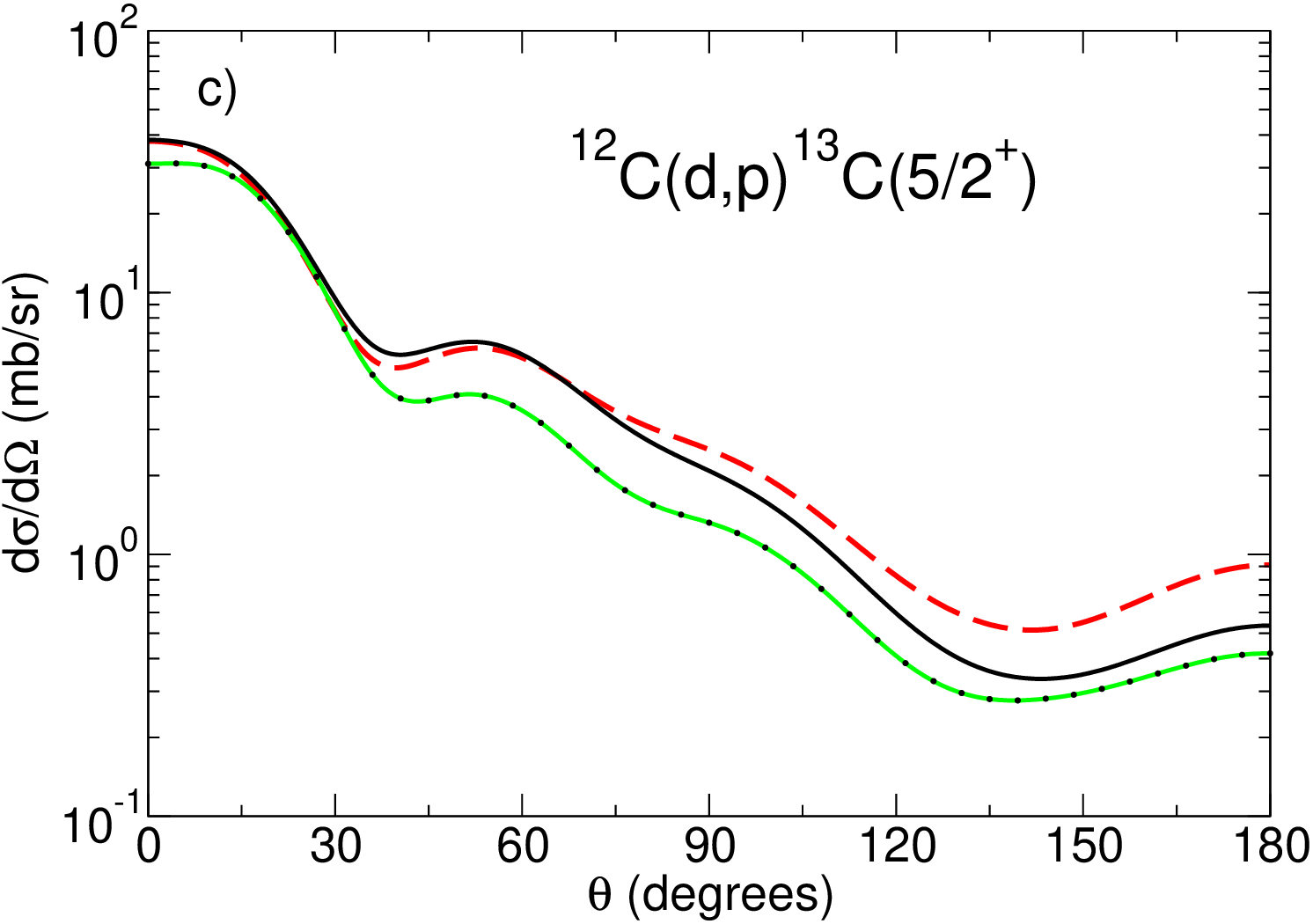}
\includegraphics[scale=0.30]{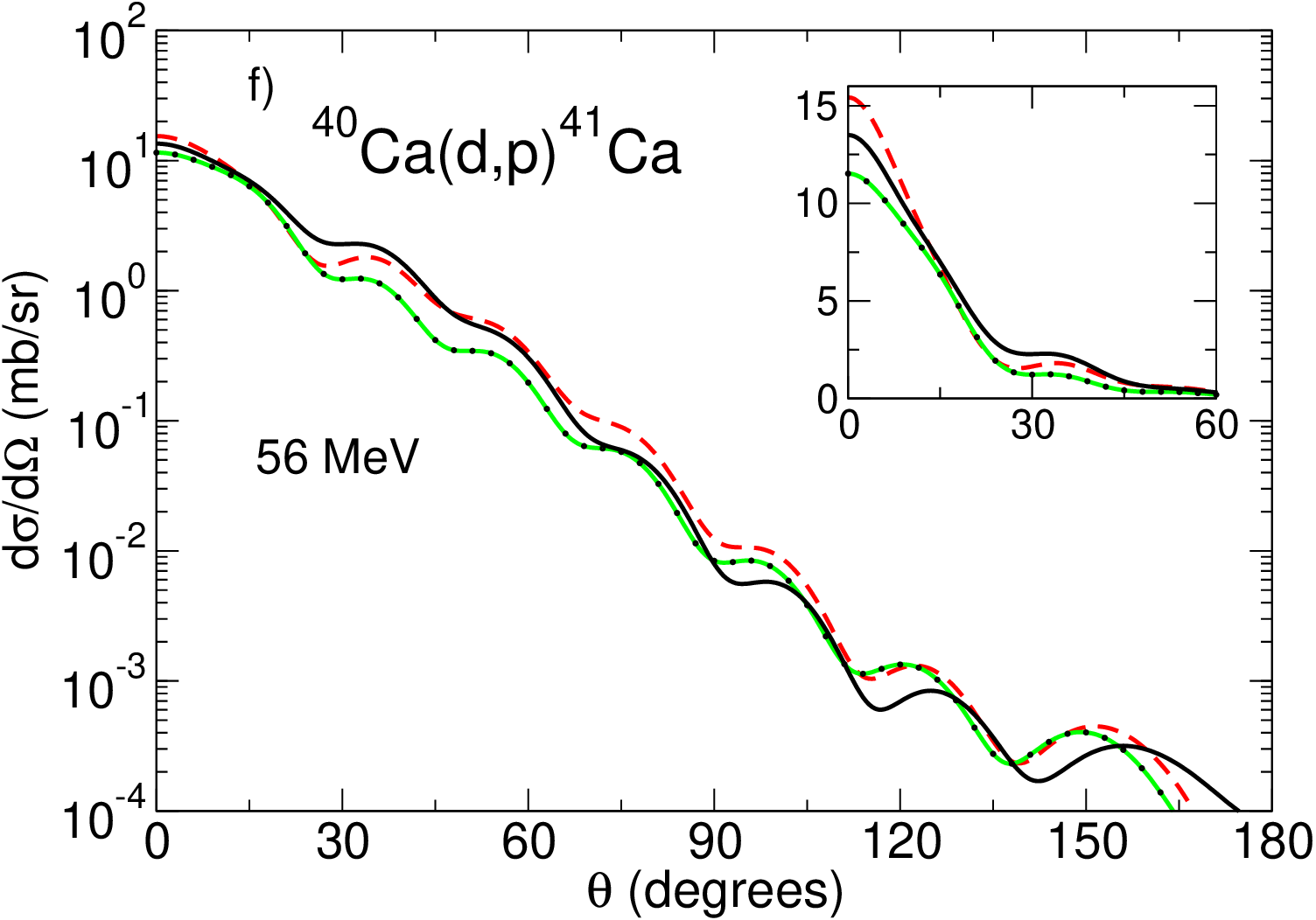}
\caption{Cross sections of the $^{12}$C($d,p)^{13}$C reaction  populating the lowest $1/2^-$ ($a$), $1/2^+$ ($b$) and $5/2^+$ ($c$) states and cross sections of the $^{40}$Ca($d,p)^{41}$Ca  reaction at 10 ($d$), 20 ($e$) and 56 ($f$) MeV calculated with effective local potentials ${\tilde U}^{\rm eff}_{NA}$ and induced three-body terms $\Delta U_1$ and $\Delta U_2$ without (dashed lines) and with (dots on  solid lines) Perey factors in comparison with the LECDCC calculations (solid line).
}
\label{fig:xsecs}
\end{figure}

The LECDCC calculations are shown in Fig. \ref{fig:xsecs} together with the local-equivalent CDCC results  obtained with ($\sigma_{\rm all}^{\mathrm{P}}$) and without ($\sigma_{\rm all}^\mathrm{noP}$) the deuteron-channel Perey factor. In all cases  they are close to $\sigma_{\rm all}$. The deviation in the first peak is 3-7$\%$ for $E_d \leq 30$ MeV but can reach 20$\%$ at 56 MeV. This is an encouraging result suggesting   that next-to-leading order LECDCC may be similar to $\sigma_{\rm all}^{\mathrm{noP}}$, for deuteron energies below 30 MeV, the typical value for transfer reaction experiments. 
 
\section {Conclusion}

Treating exactly the nonlocal nucleon-target interactions in the $(d,p)$ reactions within the CDCC has not yet been done because no methods to calculate the relevant matrix elements have been developed. We have proposed the CDCC model of $(d,p)$ reactions that includes the nonlocality of the nucleon optical potentials via the Perey-effect. This model assumes that a nonlocal optical two-body model of the Perey-Buck type has an approximately equivalent representation in term of a local optical potential  supplemented by a velocity-dependent part. Including velocity-dependent potentials in the three-body $A+p+n$ model leads to factorization of the total wave function via a product of two nucleon Perey-factors and a solution of a Schr\"odinger equation that does not contain velocity-dependence on coordinate $\ve{R}$.

It is well known that in the leading order the local-equivalent potentials $U^0_{loc}$ are found from the transcendental equation (Eq. \ref{Uloc0}). The local potentials used in the CDCC differ from $U^0_{loc}$ and this difference can affect the maxima of CDCC cross sections by 2-10$\%$. In addition, velocity-dependence generates a three-body potential whose contribution is proportional to $1/A$, so for most nuclei it can be neglected.

The Perey-effect can be easily incorporated  into available CDCC codes as long as the remnant term in the $(d,p)$ amplitude is not included. In this case, due to the short-range of the interaction $V_{np}$, one can simply multiply the overlap function for the transferred nucleon %instead of deuteron distorted wave 
by nucleon Perey factors. Their influence on CDCC cross sections is largest for reactions where the contribution from the nuclear interior is not suppressed. Also, their importance increases with the incoming deuteron energy. While for energies typical for $(d,p)$ reaction experiments the Perey-effect could affect the main peak, used for spectroscopic factors determination, by 3-7$\%$, for higher energies it can reach up to 20$\%$, and thus must be taken into account when extracting spectroscopic factors. On the other hand, for  peripheral reactions, in particular at low incident energies, the Perey-effect can be neglected. In this case nonlocality can be mainly described through the modification of $U^0_{loc}$. 

We have checked that the CDCC with velocity-dependent potentials without the Perey-effect gives cross sections similar to those obtained in the CDCC, based on nonlocal optical potentials, treated within the leading-order local-energy approximation. This gives an encouraging indication to the possibility that including the Perey-effect within CDCC, as has been developed here, may represent a full nonlocal problem in a satisfactory way. 

The estimates shown here can give an idea of the uncertainties, arising due to nonlocality, in spectroscopic factors and asymptotic normalization coefficients extracted using CDCC. If they are found by comparing the cross sections in the main peak only then these uncertainties could be comparable with those of experimental measurements. However, in some cases the introduction of the Perey-effect affects the slope of the angular distributions as well. More research is needed for such cases.  

Finally,  accounting for nonlocality in the CDCC, proposed in this paper, is based on analytical representation (\ref{PB}) of the nonlocal potentials. The optical potentials obtained in many-body approaches, in general, do not have this form \cite{Rot17}. Approximating such potentials by analytical functions of $\ve{r}-\ve{r}'$ and $(\ve{r}+\ve{r}')/2$ would make possible  the approximate treatment of their nonlocality along the lines suggested here. Otherwise new methods of the CDCC matrix elements calculation should be developed.

\section*{Acknowledgements}
This work was supported by the United Kingdom Science and Technology Facilities Council (STFC) under Grant No.  ST/P005314/1. M.G.-R. acknowledges funding from the Spanish Government under project No. FIS2014-53448-C2-1-P and FIS2017-88410-P and by the European Unions Horizon 2020 research  and  innovation  program  under  grant  agreement  No. 654002. He also acknowledges partial support from the Nuclear.Theory.Vision$@$UK visitors program.

\end{document}